\documentclass[]{spie}  

\usepackage{amsmath,amsfonts,amssymb}
\usepackage{graphicx}
\usepackage[colorlinks=true, allcolors=blue]{hyperref}
\usepackage[utf8]{inputenc}

\title{Preliminary results of the pixel characterization for the Crystal Eye, a new X and $\gamma$-ray satellite detector for multi-messenger astronomy}

\author[a,b]{F.C.T. Barbato}
\author[c]{G. Barbarino}
\author[c]{A. Boiano}
\author[c]{A. Vanzanella}
\author[c,d]{F. Garufi}
\author[c,d]{F. Guarino}
\author[e]{F. Renno}
\author[e]{S. Papa}
\author[e]{R.Guida}
\author[c,d]{F. Di Capua}

\affil[a]{Gran Sasso Science Institute (GSSI),  L'Aquila, Italy}
\affil[b]{Istituto Nazionale di Fisica Nucleare, Laboratori Nazionali del Gran Sasso, Assergi, Italy}
\affil[c]{Istituto Nazionale di Fisica Nucleare, Sezione di Napoli,  Napoli, Italy}
\affil[d]{Università degli Studi di Napoli Federico II, Dipartimento di Fisica "E. Pancini",  Napoli, Italy}
\affil[e]{Università degli Studi di Napoli Federico II, Dipartimento di Ingegneria Industriale,  Napoli, Italy}

\authorinfo{Further author information: (Send correspondence to Felicia Barbato)\\Felicia Barbato : E-mail: felicia.barbato@gssi.it}

\begin{document} 
\maketitle

\begin{abstract}
With the observation of the gravitational wave event of August 17th 2017 the multi-messenger astronomy era has definitely begun. With the opening of this new panorama, it is necessary to have new instruments and a perfect coordination of the existing observatories. \\
Crystal Eye is a detector aimed at the exploration of the electromagnetic counterpart of the gravitational waves. Such events generated by neutron stars’ mergers are associated with gamma-ray bursts (GRB). At present, there are few instruments in orbit able to detect photons in the energy range going from tens of keV to few MeV. These instruments belong to two different old observation concepts: the all sky monitors (ASM) and the telescopes. The detector we propose is a crossover technology, the Crystal Eye: a wide field of view observatory in the energy range from 10 keV to 10 MeV with a pixelated structure. A pathfinder will be launched with Space RIDER in 2022.\\ 
We here present the preliminary results of the characterization of the first pixel.
\end{abstract}


\section{INTRODUCTION}
\label{sec:intro}  

\begin{figure}[h!]
		\centering
		\includegraphics[scale=.6]{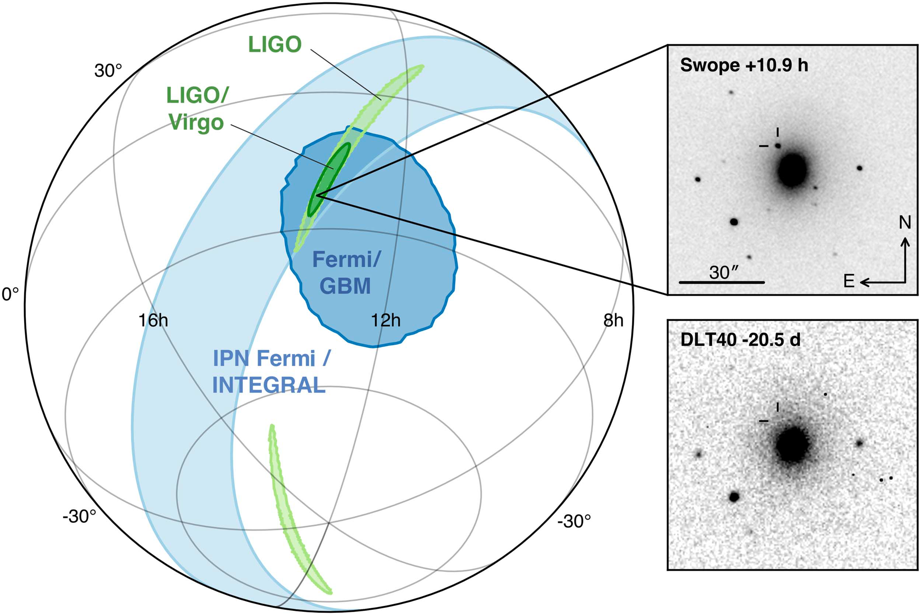} 
		\label{fig:Fermi-GRB}\caption{Final localization of the source that generated both the GW170817 and the GRB170817A.}
	\end{figure}

Multimessenger astrophysics is a new way of exploring the Universe, powered by globally coordinated observations of cosmic rays, neutrinos, gravitational waves, and electromagnetic radiation across a broad range of wavelengths.\\
With the observation of the gravitational wave event GW170817 of August 17th 2017 and then with those of the extragalactic neutrino of September 22nd, the multimessenger astrophysics era has definitely begun.\\
A long-standing astrophysical paradigm is that collisions, or mergers, of two neutron stars create highly relativistic and collimated jets that power gamma-ray bursts of short duration. The observational support for this model, however, was only indirect until August 2017, when the electromagnetic counterpart associated with the gravitational-wave event GW170817 was observed.
The Fermi-GBM and INTEGRAL instruments, about 2s later observed independently a short gamma-ray burst, GRB 170817A in the same sky region\cite{abbott2017multi}\cite{abbott2017gw170817}\cite{abbott2017gravitational}\cite{valenti2017discovery}\cite{goldstein2017ordinary}\cite{troja2017x} , see FIG. \ref{fig:Fermi-GRB} 

Deaspite all the excitement about this unique event, in 2019 a new neutron stars merger was detected by LIGO-Virgo collaboration GW190425\cite{abbott2020gw190425}. For this event INTEGRAL claimed the detection of a faint GRB \cite{pozanenko2019observation}. Nevertheless this event remains very debated since no other experiments where able to detect the electromagnetic counterpart.\\
This conflict between the two events of neutron star mergers, GW170817 and GW190425, can be interpreted as even more exciting since it leaves the questions about the origin of s-GRBs still unanswered.\\
In this framework it is clear that we need new observatories as well as new observational techniques to have the chance to finally answer this question.\\
Crystal Eye detection module is our proposed solution to improve X and $\gamma$-rays detection in the energy range 10 keV - 10 MeV.

\section{Crystal Eye detection module}
\label{sec:CE}  

\begin{figure}[h!]
		\centering
		\includegraphics[scale=.28]{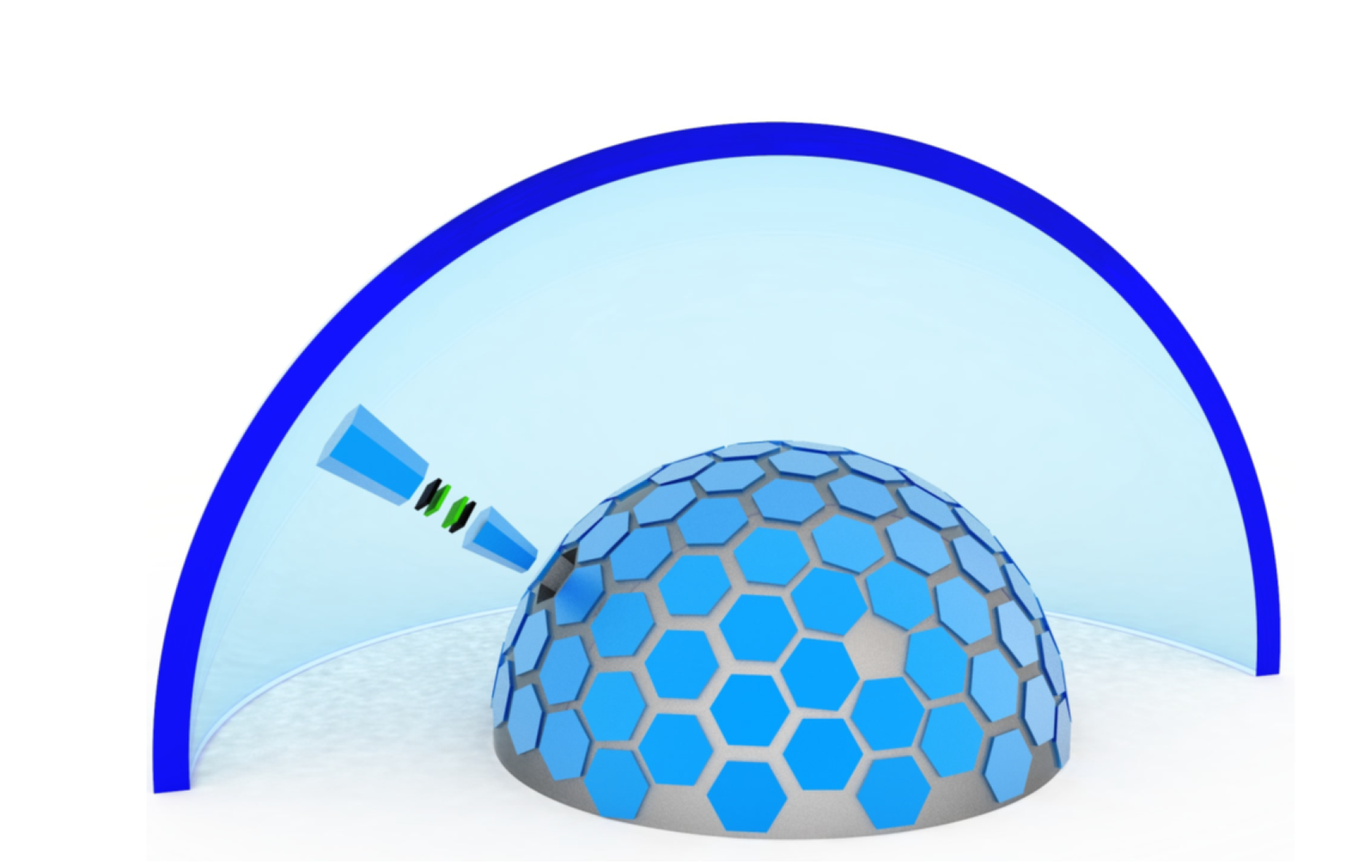} 
		\label{fig:crystaleye}\caption{Exploded picture of the selected Crystal Eye design.}
	\end{figure}

Crystal Eye X and $\gamma$-rays detection module is a new concept of all sky monitor designed to explore the energy range 10 keV - 10 MeV with a higher resolution with respect to the current all sky monitors\cite{barbato2019crystal}. \\
It is coinceived to be a crossover technology between all sky monitors and telescopes. The idea is indeed to have a semispherical detection module with a high pixel number in order to obtain a wide field of view and at the same time a better localization capability.\\
Crystal Eye is designed to fly on LEO orbits and to be very compact in order to be eventually portable by human flights on space stations. Its diameter is indeed only 30 cm. With this constraints, three possible designs have been studied. The common configuration consists in a shell structure made by two layers of LYSO crystals, each read by a SiPM array, and an anticoincidence module for charged particle discrimination.\\ Thanks to a selection method based on analytic hierarchy process \cite{renno2020ahp} , the selected configuration for the Crystal Eye forsees 112 pixels each made by two LYSO crystals with a truncated pyramidal shape with hexagonal bases, see FIG. \ref{fig:crystaleye}, namely UP pixel and DOWN pixel.
	 \\
From the figure, it is evident that the high coverage ensured by the hexagonal pyramid shape of the LYSO crystals lead to a reasonable high detection area ($\sim 800 \: cm^2$). These features are now possible thanks to the evolution of technology. SiPMs experienced in the last decade an incredible progress and it is now the right time to exploit also in space experiments their incredible detection features, such as the excellent photon detection capability, the high dynamic range, the negligible power consumption and the insensitivity to magnetic fields.\\
The outer shell is made by a thin layer (0.5cm) of plastic scintillator, read by SiPMs, that works as anticoincidence module for charged particle discrimination.\\
The semisphere shape guarantees the observation of the sky with a wide field of view (2$\pi$ locally) and the full scan of the sky along the orbit (4$\pi$ in $\sim 90$ minutes).\\
This shape has actually several aims, it indeed guarantees:
	\begin{itemize}
		\item \textbf{symmetry}, all the directions are equivalent; 
		\item \textbf{thermal protection}, the SiPMs are shielded by LYSO crystals from the external environment ensuring a thermal stabilization in the SiPM housing layer;
		\item \textbf{radiation hardness}, the SiPM housing layer is shielded by 4cm of LYSO on the top and 3cm of LYSO on the bottom, strongly reducing the radiation damage probability of the sensors; 
	\end{itemize}

\subsection{The pixel}
\label{sec:pixel}  
\begin{figure}[h!]
		\centering
		\includegraphics[scale=.5]{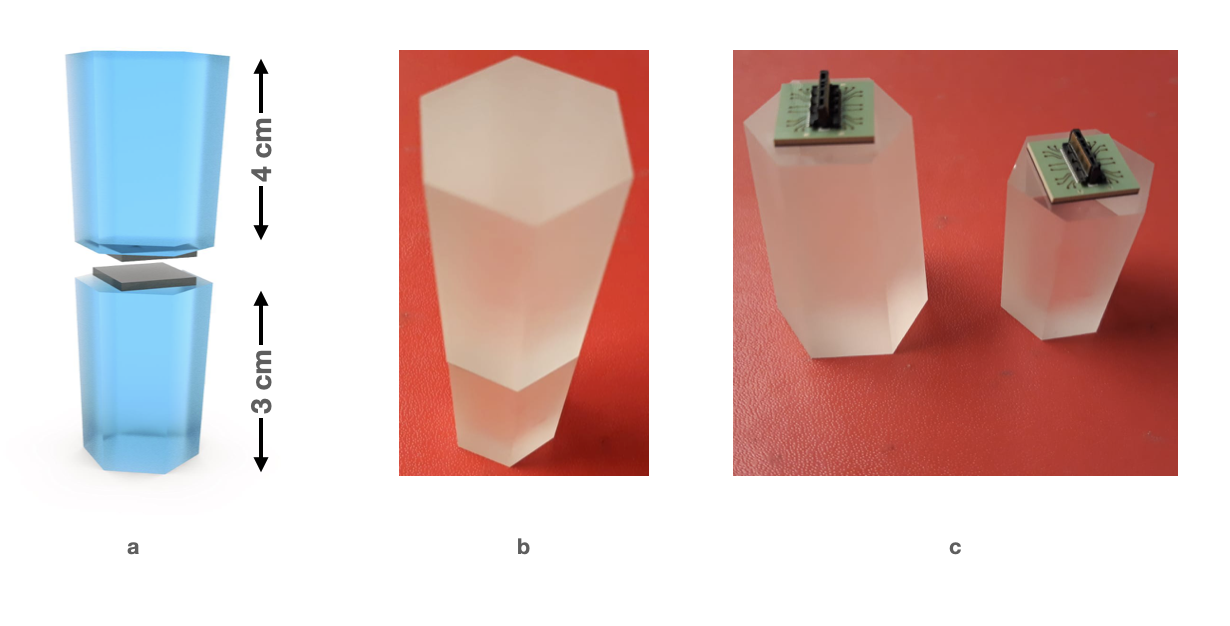}
		\label{fig:pixel}\caption{\textbf{a)} Design of a pixel of the Crystal Eye module. \textbf{b)} Picture of the LYSO pixel. \textbf{c)} Picture of the UP and DOWN pixels with the SiPM arrays.}
	\end{figure}

The Crystal Eye pixel will be made by hexagonal trunks of pyramid of LYSO crystals. Each pixel is composed by two parts, a UP pixel and a DOWN pixel, designed to be one the continuum of the other. Between the two pyramids an additional layer is interleaved. This is dedicated to the housing of SiPMs and their front end electronics, see FIG. \ref{fig:pixel}.
As already said, the shape of the crystals has been designed in order to ensure the maximum covering of the semispherical surface.\\
The LYSO scintillator has been chosen for several reasons:
\begin{itemize}
	\item the gamma ray absorption probability is higher than in other materials;
	\item the time response is very fast ($\sim$36 ns);
	\item the light yield is very high ($\sim 33\gamma$/keV);
	\item the emission spectrum can be used for self-calibration of the sensors\cite{alva2018understanding} .
\end{itemize}

\subsection{The Crystal Eye method}
\label{sec:CE_method}  
The Crystal Eye detection method is based on the readout of the charge distribution along the detector surface.\\
The symmetry of the detector and the double layer structure with the anticoincidence module are necessary to allow source localization and particle identification.
	\begin{figure}
		\centering
		\includegraphics[scale=.6]{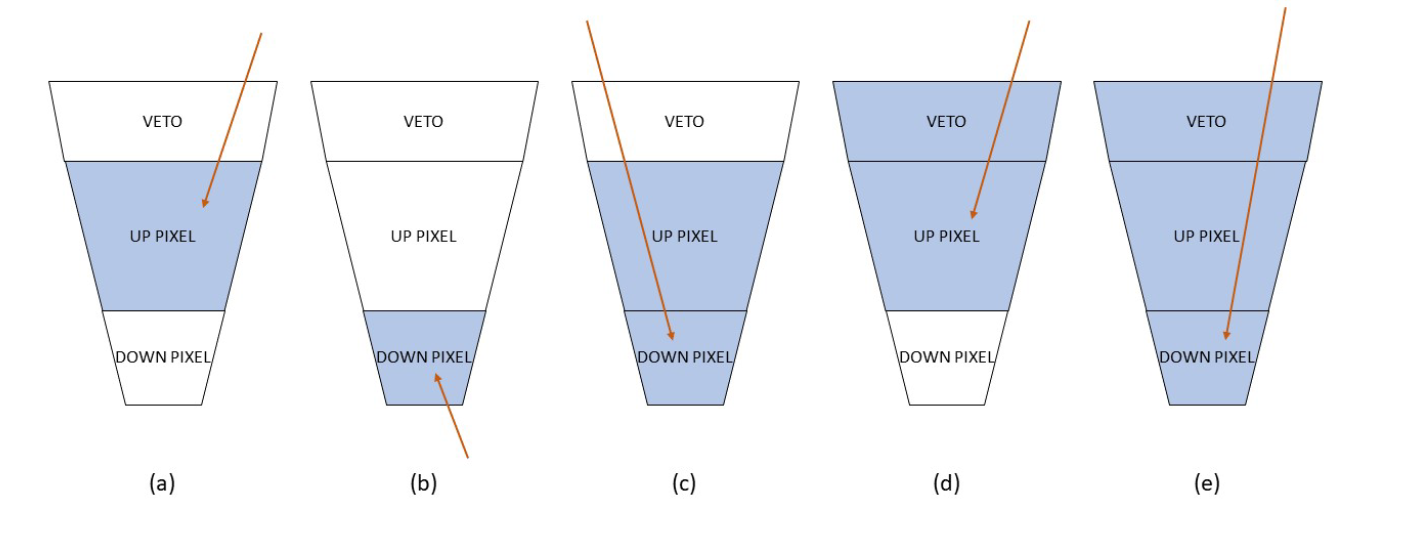}
		\label{fig:lettura}\caption{Crystal Eye possible triggers.}
	\end{figure}
The size of the LYSO pyramids is optimized to absorb gamma rays with energy below 1MeV in the UP pixel with 85\% probability.\\
In FIG. \ref{fig:lettura} the possible trigger configuration of the Crystal Eye are shown for:

	\begin{enumerate}
		\item[\textbf{a.}] Down-going gamma ray with E$<$ 1MeV;
		\item[\textbf{b.}] Up-going gamma ray with E$<$ 1MeV;
		\item[\textbf{c.}] Gamma ray with E$>$ 1MeV;
		\item[\textbf{d.}] Low enegy charged particle;
		\item[\textbf{e.}] High enegy charged particle.
	\end{enumerate}

In FIG. \ref{fig:lettura}c, a double possibility is present. The gamma ray can be indeed either down-going and up-going. Therefore UP and DOWN pixel heights were differentiated in order to create an asymmetry in the signal.\\
FIG. \ref{fig:lettura}a and c represent the cases of good triggers for GRB detection.\\
By reading the charge distribution in case of good triggers, a single Crystal Eye module will improve the localization capability of about three times with respect to Fermi-GBM. A constellation of three modules will ensure a constant full sky monitor as well as a further improved localization capability thanks to the triangulation of the signals coming from the three modules in addition to the charge distribution readout. Moreover, having three identical modules in orbit means also a sufficiently high detection surface as well as a sufficient distance to explore an eventual submillisecond time structure of the observed GRB. 

\section{Preliminary measurements on LYSO crystals}
\label{sec:measurements}  
\begin{figure}[h!]
		\centering
		\includegraphics[scale=.3]{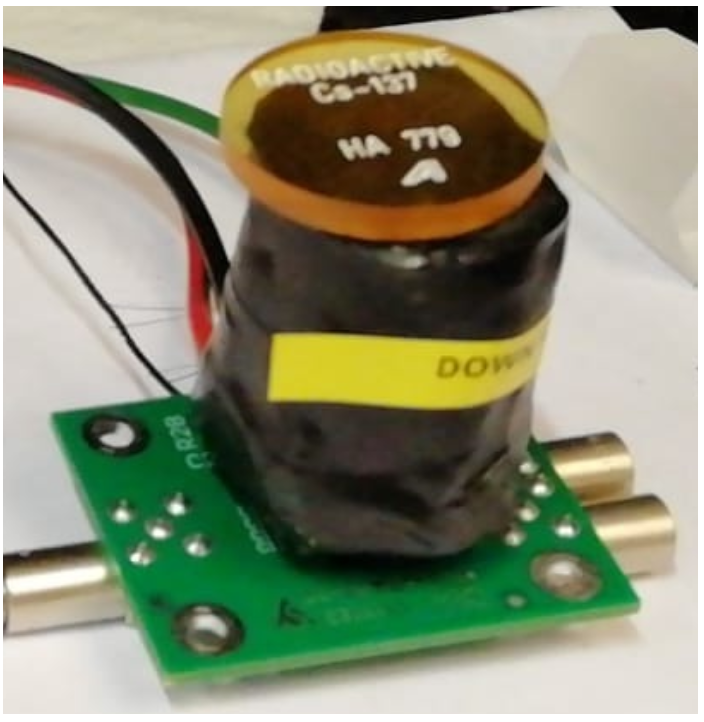} 
		\label{fig:pixel_misura}\caption{DOWN Epic crystal packed during measurements.}
	\end{figure}

In 2022 a pathfinder made by 4 pixels is expected to fly onboard of the Space RIDER vehicle by ESA for a 60 days mission\cite{renno2020systems}.\\
In view of this mission, we ordered LYSO pyramids from two companies: Epic Crystals and OST Photonics. The pyramids surfaces were all grounded except for those interfacing the SiPM array. All of them were first wrapped by using thin teflon tape 0.2mm thick and then were obscured by means of 4mm thick black heat shrink tube, see FIG \ref{fig:pixel_misura}.

\subsection{Front-end electronics and DAQ}
\label{sec:elettronica}
\begin{figure}[h!]
		\centering
		\includegraphics[scale=.5]{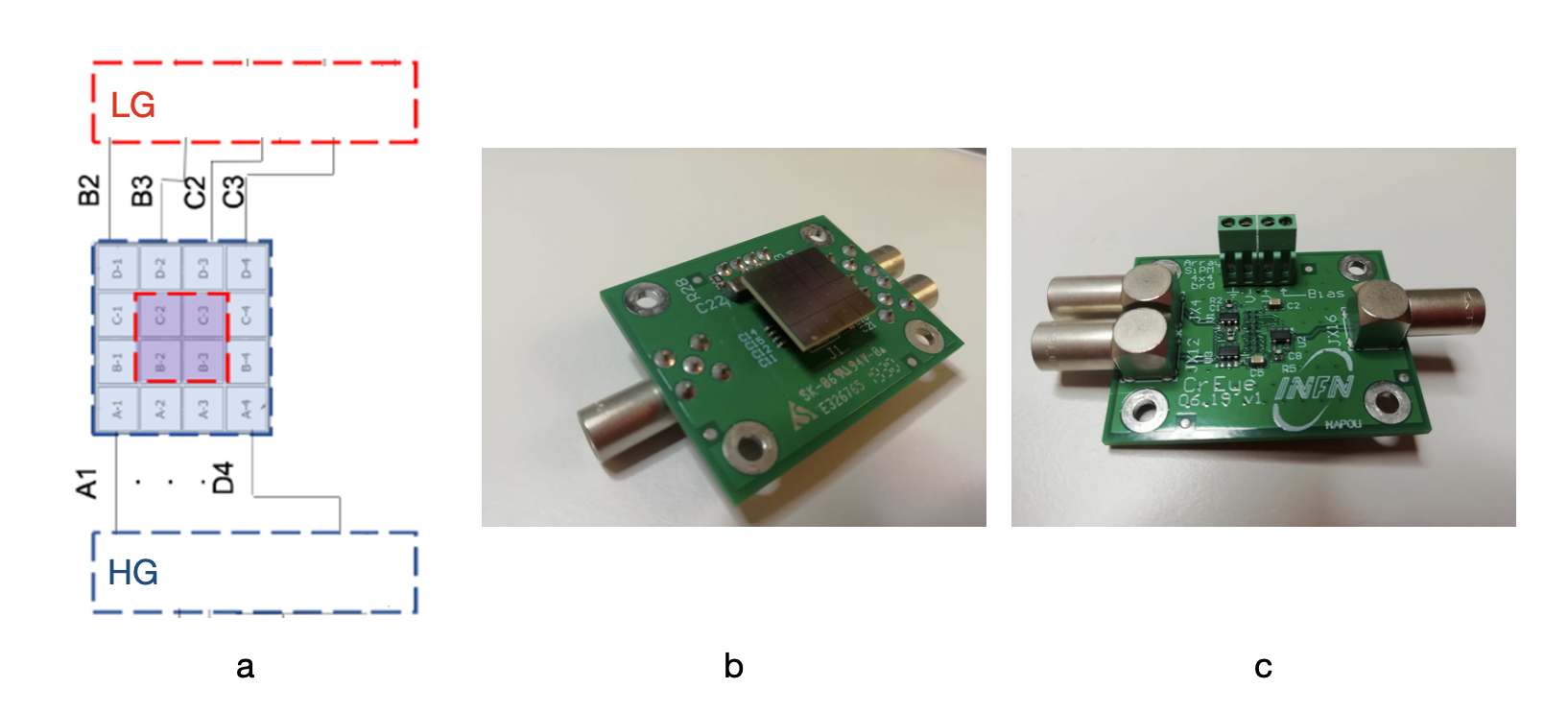} 
		\label{fig:frontend}\caption{a) Front-end electronics concept. b) Front-end board with SiPM array (top view). c) Front-end board (bottom view).}
	\end{figure}

The LYSO Crystals are interfaced with 4$\times$4 MPPC arrays by Hamamatsu S13361-3050AE-04\cite{hamamatsu} . This particular array was chosen because of its size that well fit the UP and DOWN pixel interface surfaces as well as having 16 SiPMs means both having redoundancy of the sensors and having the possibility to split the SiPM outputs in two groups in order to have a high gain (HG) channel and a low gain (LG) one.\\
The front-end electronics designed at the INFN-Sezione di Napoli provides the LG-channel by summing up the signal of the four central SiPMs and a HG-channel by summing up the signals of all the 16 SiPMs, see FIG. \ref{fig:frontend}a
	\\
The LG-channel is enpowered to read the signals of gamma rays with E $>$ 1MeV, while the HG-channel is for gamma rays with E $<$ 1MeV. Since the output signal is read by the A1702 CAEN DAQ board, which is based on the WeeROC CITIROC ASIC, the two gain channels are necessary in order to avoid the electronics saturation as well as for exploiting all the ADC range for both energy ranges.\\

\subsection{LYSO pixel by Epic Crystals}
\label{sec:Epic}
\begin{figure}[h!]
		\centering
		\includegraphics[scale=.4]{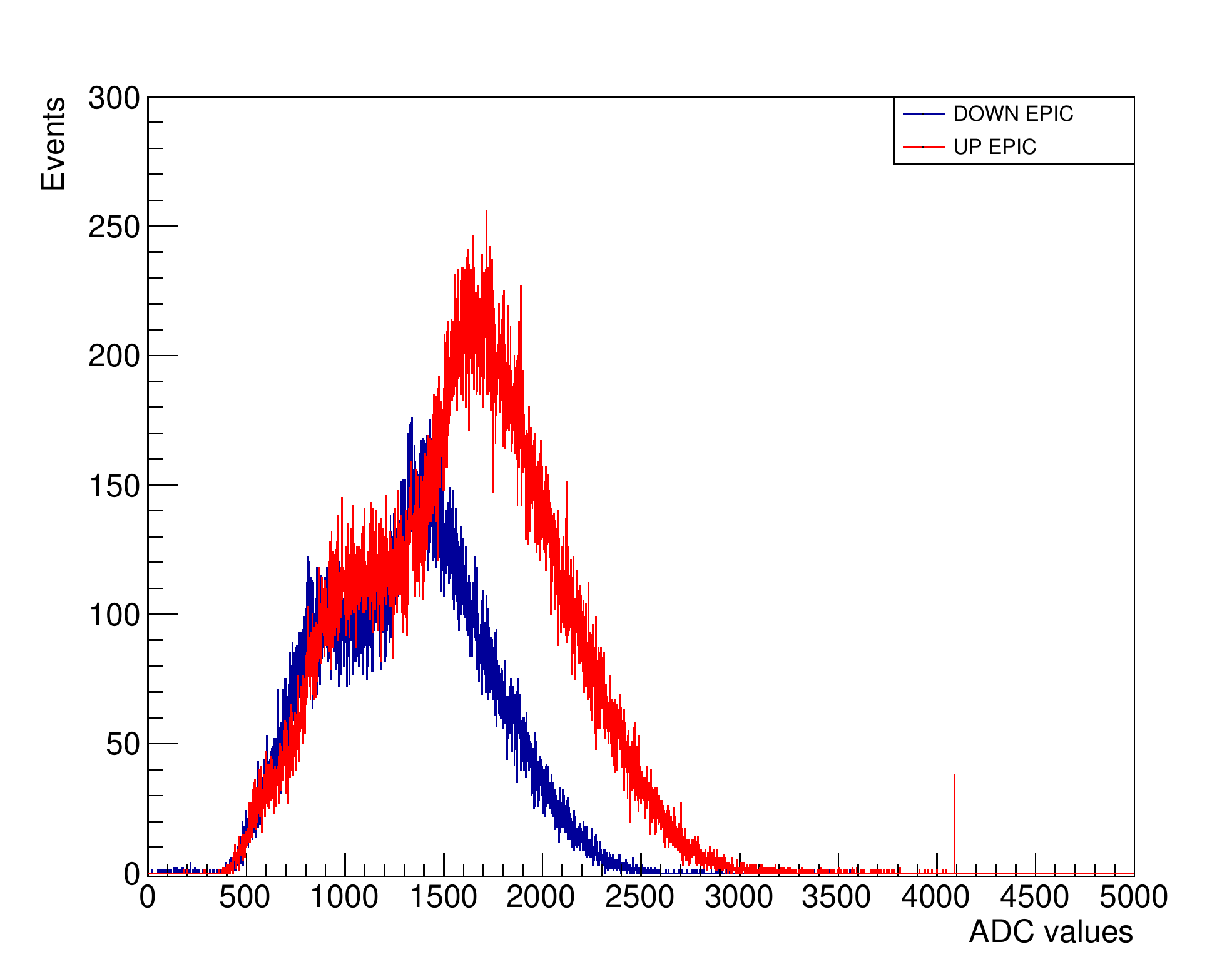} 
		\label{epic_updown}\caption{Emission spectra of UP and DOWN pixels manufactured by Epic Crystals}
		\end{figure}

We first measured the LYSO emission spectrum of both UP and DOWN pixels manufactured by Epic Crystals, see FIG. \ref{epic_updown}.

Then we tested both by means of radioactive sources: Co-60, Ba-133 and Cs-137.

	\begin{figure}[h!]
		\centering
\includegraphics[scale=.62]{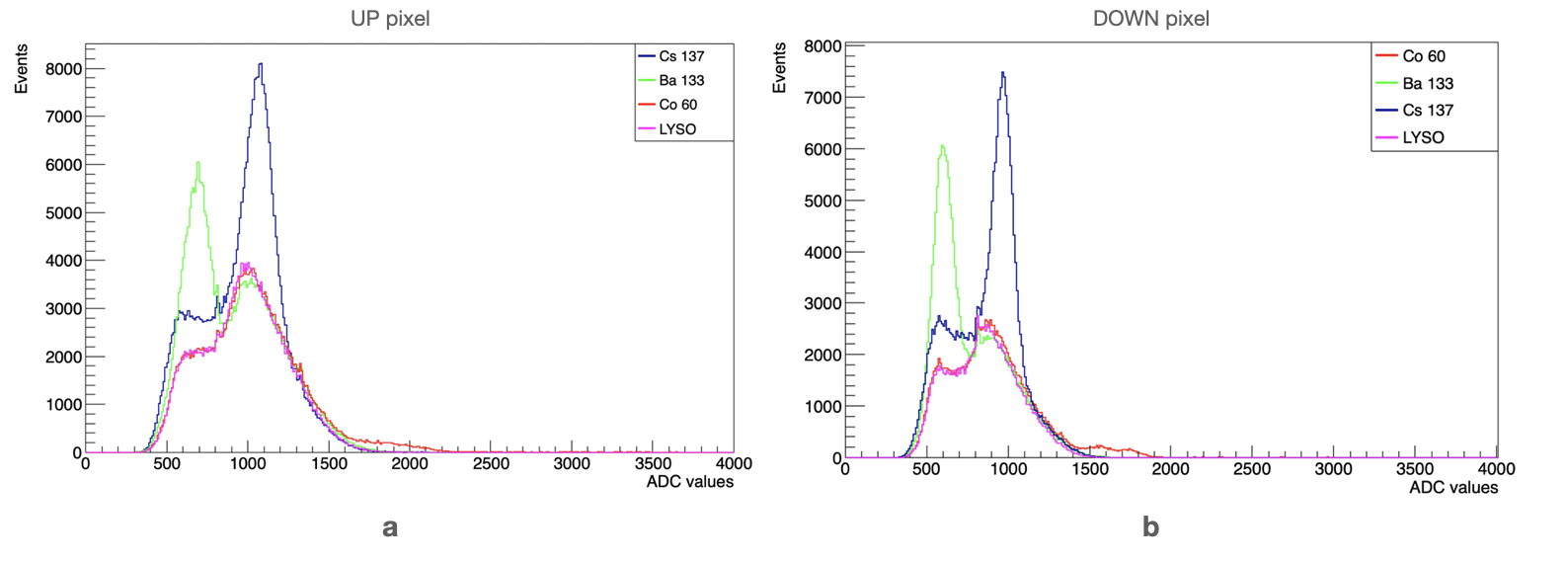}
		\label{epic_updown_sorgenti}\caption{Absorption spectrum of radioactive sources in LYSO pixels by Epic Crystals. a) UP pixel, b) DOWN pixel HG-channel.}
		\end{figure}

Based on the design of the front end electronics, the ADC range in the HG channel is expected to be up to 1 MeV. In FIG. \ref{epic_updown_sorgenti}, it is evident that the signal is not optimizing the ADC range since the Co-60 signal that is around 1 MeV is located below 2000ADC counts for both pixels. Energy resolution of the peaks in the UP pixel results to be 14\%, while it is 13\% in DOWN pixel.

\subsection{LYSO pixel by OST Photonics}
\label{sec:OST}
\begin{figure}[h!]
		\centering
\includegraphics[scale=.5]{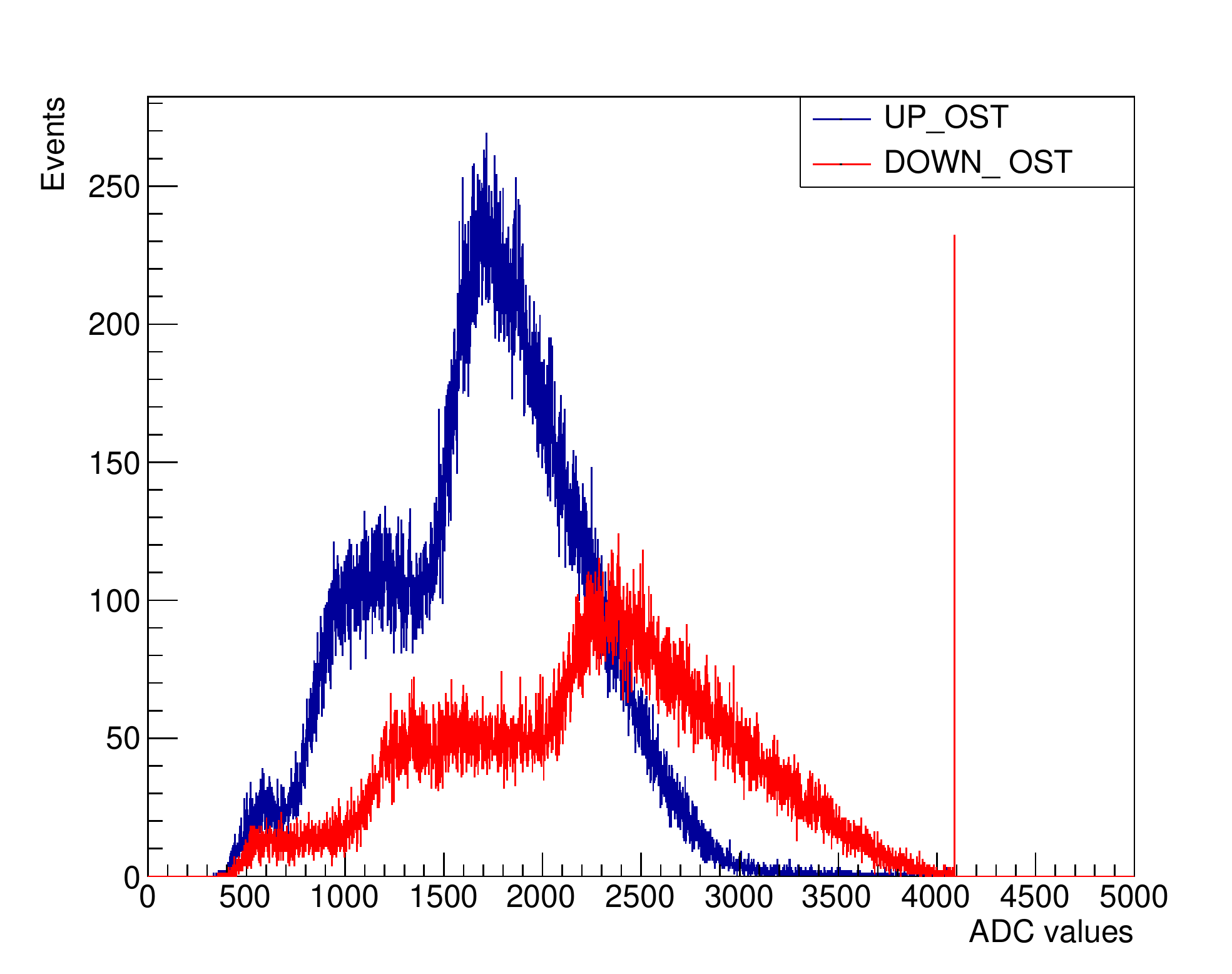} 
		\label{ost_updown}\caption{Emission spectra of UP and DOWN pixels manufactured by OST Photonics}
		\end{figure}

Also in this case, we first measured the LYSO emission spectrum of both UP and DOWN pixels manufactured by OST Photonics, see FIG. \ref{ost_updown}.

Then we tested both by means of radioactive sources: Co-60, Ba-133 and Cs-137.

	\begin{figure}[h!]
		\centering
\includegraphics[scale=.62]{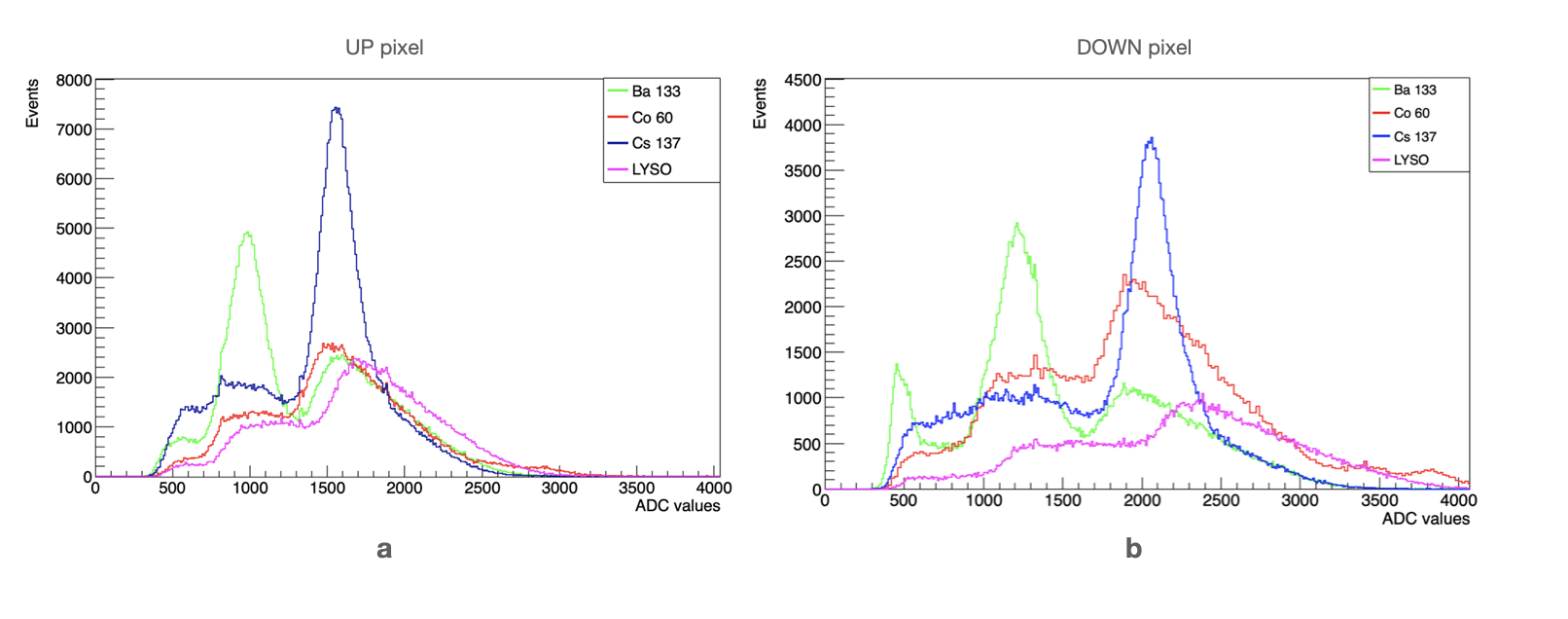}
		\label{ost_updown_sorgenti}\caption{Absorption spectrum of radioactive sources in LYSO pixels by OST Photonics. a) UP pixel, b) DOWN pixel HG-channel.}
		\end{figure}

\begin{figure}[h!]
		\centering
		\includegraphics[scale=.5]{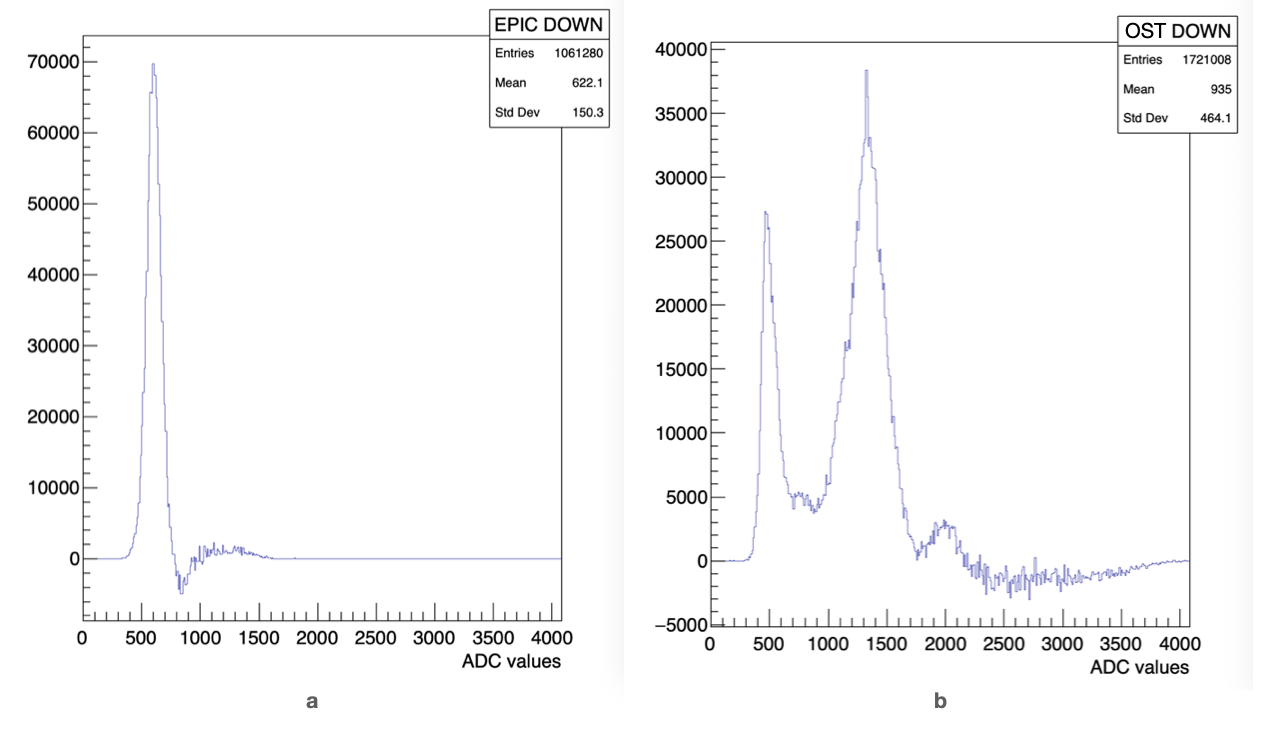} 
		\label{bario}\caption{Barium normalized spectrum. a) DOWN pixel by Epic Crystals, b) DOWN pixel by OST Photonics}
	\end{figure}

As evident from FIG. \ref{ost_updown_sorgenti}, in this case the signal is optimizing the ADC range since the Co-60 signal is located above 2000ADC counts for the UP pixel and above 3000 ADC counts for the DOWN pixel. Thanks to this well optimized configuration it is possible a better discrimination of the absorption peaks.\\
In particular, for the Ba-133 spectrum in OST Photonics pixel in addition to the 356 keV peak which shows the highest branching ratio, the peak corresponding to 81 keV is also well visible, see \ref{bario}.
Thanks to this very good result, for the OST Photonics pixel we measured the spectra also in the LG channel.

	\begin{figure}[h!]
		\centering
		\includegraphics[scale=.5]{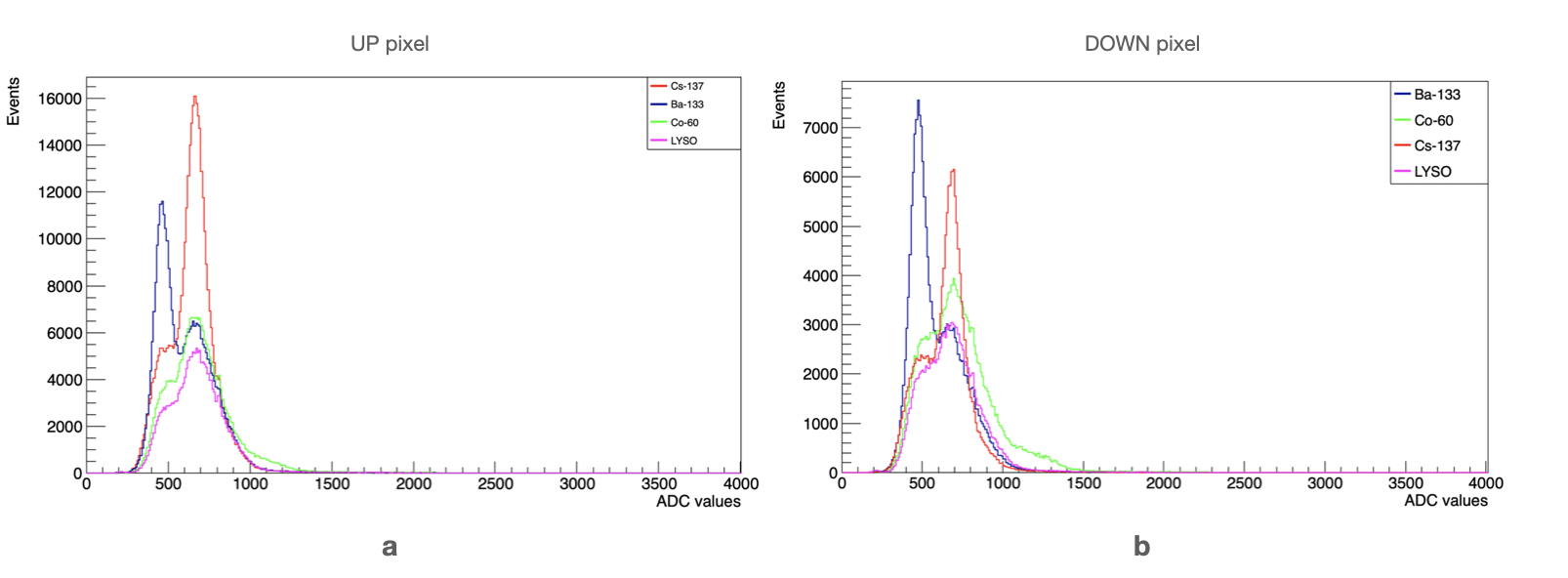} 
		\label{ost_LG_sorgenti}\caption{Absorption spectrum of radioactive sources in LYSO pixels by OST Photonics. a) UP pixel, b) DOWN pixel LG-channel.}
	\end{figure}

The LG channel of the crystals manufactured by OST Photonics shows a behaviour close to that of the pixels by Epic Crystals. As a perfect example, in this case the 81keV peak of the Ba-133 is not anymore visible. The fact that in the LG channel all the spectra are compressed down the ADC range is exactly what we expected, so the ADC will range up to $\sim$4 MeV for each pixel.\\

\section{Conclusions}
The Crystal Eye X and gamma-rays detector is a new concept design of all sky monitor. With this modern design it is possible to achieve a source localization capability up to three times better than Fermi-GBM. The use of a constellation made by three modules can further improve the localization capability.\\
In 2022 a technological pathfinder will be launched for a 60 days mission onboard of the Space RIDER vehicle by ESA. In view of the launch, we bought and tested a sample of the Crystal Eye pixel by two different companies: OST Photonics and Epic Crystals.\\
A first set of measurements was done on these samples. From this first dataset it seems that the pixels manufactured by OST Photonics have a better sensitivity with respect to those by Epic Crystals. Nevertheless, since the energy resolution of the visible peaks of those by Epic Crystals is systematically 2\% better than those of OST Photonics, we suppose that this difference could be due to a lower amount of scintillation photons detected by the SiPM array and so to a worse optical coupling among the SiPM-arrays and the crystals. Further measurements will be done by repeating the optical coupling of the Epic Crystals sample in order to confirm this hypotesis.

\acknowledgments 
 The Crystal Eye R\&D is funded by UNINA and Intesa San Paolo in the framework of the \textit{Bando STAR2018 - L1 Junior Principal Investigator}.\\
 Authors want to thank Dr. Pierluigi Casolaro for his support in analyzing data and Milena Genzini for the amazing logo of the experiment.

\bibliography{Pixel_bib} 
\bibliographystyle{spiebib} 

\end{document}